\documentclass[12pt]{article}

\usepackage{graphicx}
\begin{document}

\begin{center}
{\bf Dyonic black holes in framework of Born$-$Infeld-type electrodynamics } \\
\vspace{5mm} S. I. Kruglov
\footnote{E-mail: serguei.krouglov@utoronto.ca}
\underline{}
\vspace{3mm}

\textit{ Department of Physics, University of Toronto, \\60 St. Georges St.,
Toronto, ON M5S 1A7, Canada\\
Department of Chemical and Physical Sciences,\\ University of Toronto Mississauga,\\
3359 Mississauga Rd. N., Mississauga, ON L5L 1C6, Canada}
\vspace{5mm}
\end{center}
\begin{abstract}
Dyonic black hole solutions with spherically symmetric configurations within general relativity are investigated where the source of the gravitational field is Born$-$Infeld-type electrodynamics. Corrections to Coulomb's law and Reissner$-$Nordstr\"{o}m solutions are obtained. From principles of causality and unitarity we find the restriction on electric and magnetic fields. The Hawking temperature and the heat capacity of black holes are obtained. At some event horizons second-order phase transitions take place.
\end{abstract}

\section{Introduction}

In low-energy string theory static and spherically symmetric charged black hole (BH) solutions were found \cite{Garfinkle}. Dilaton and axion are presented in the solution. The electric magnetic duality of dyonic (with electric and magnetic charges) BH solution in the string theory was discussed in \cite{Shapere}, \cite{Sen}. Some aspects of dyonic BH solutions in string theory were studied in \cite{Mignemi}-\cite{Jatkar}. BH dyonic solutions were considered in the theory of supergravity \cite{Chamseddine}-\cite{Meessen}.
 The Hall conductivity due to dyonic BH solutions in the framework of AdS/CFT correspondence was considered in \cite{Hartnoll}. The results are in agreement with the hydrodynamic analysis.
Anti de Sitter (AdS) BH with electric and magnetic charges describes the Nernst effect \cite{Hartnoll1}. It was shown, using the AdS/CFT correspondence, that dyonic BHs in AdS spacetime are dual to stationary solutions of the relativistic magnetohydrodynamics equations \cite{Caldarelli}.
The superconductivity was investigated by considering AdS/CFT duality \cite{Albash}. Thermodynamics of dyonic BH, phase transitions and critical phenomena were studied in \cite{Dutta}, \cite{Hendi}. All these shows the importance of dyonic BH solutions.

Here, we obtain and investigate new dyonic BH solutions with the nonlinear Born$-$Infeld-type electrodynamics.
Nonlinear electrodynamics (NED) can solve problems of singularities in the center of particles and infinite self-energy at the classical level. First NED was proposed by Born and Infeld (BI) \cite{Born} which can solve problems of singularities. Quantum electrodynamics due to loop corrections also produces NED \cite{Heisenberg}. Some NEDs were proposed in \cite{Soleng}-\cite{Kruglov1} where singularities are absent. In \cite{Pellicer}-\cite{Quiros} NED's coupled to general relativity (GR) were investigated. Thermodynamics of BH was studied in \cite{Hendi}-\cite{Kruglov2}. Electric and magnetic charged BHs were investigated in the framework of NED in \cite{Yajima}-\cite{Kruglov3}. It was shown that phase transitions occur in some models of BH.
NEDs coupled to GR also describe inflation and current acceleration of the universe  \cite{Garcia}-\cite{Kruglov4}.

The paper has the structure as follows. The model and principles of causality and unitarity of Born$-$Infeld-type electrodynamics are studied in Sec. 2. The bounds on electromagnetic fields were obtained. The dyonic solution for BH and corrections to Coulomb's law and RN solutions are found in Sec. 3. It was shown that there can be naked singularities, extremal BH solutions and BH solutions with two horizons at some model parameters. Thermodynamics and phase transitions of BH are investigated in Sec. 4. The Hawking temperature and heat capacity of BH are calculated. We demonstrate that phase transitions in BH can occur. In Sec. 5 we make a conclusion. In Appendix singularities are studied and we calculate the Kretschmann scalar.

The units with $c=1$, $k_B=1$, and signature $\eta=\mbox{diag}(-,+,+,+)$ are used.

\section{The model and principles of causality and unitarity}

Let us consider the Lagrangian density of BI-type electrodynamics which is a particular case of a model (at $\sigma=3/4$) proposed in \cite{Kruglov1} (see also \cite{Krug12})
\begin{equation}
{\cal L} = \frac{1}{\beta}\left[1-\left(1+\frac{4\beta{\cal F}}{3}\right)^{3/4}\right],
 \label{1}
\end{equation}
where ${\cal F}=(1/4)F^{\mu\nu}F_{\mu\nu}=(\textbf{B}^2-\textbf{E}^2)/2$ and the field strength tensor is $F_{\mu\nu}=\partial_\mu A_\nu-\partial_\nu A_\mu$. The parameter $\beta$ ($\beta>0$) has the dimension of (length)$^{4}$ and $\beta\cal F$ is dimensionless. The correspondence principle holds because at ${\cal F}\rightarrow 0$ the Lagrangian density (1) becomes the Maxwell Lagrangian density ${\cal L}_M=-{\cal F}$.
Here, we consider BI-type electrodynamics in the framework of special relativity. At the center of charges, the maximum of the electric field is finite and is $E_{max}=\sqrt{3}/\sqrt{2\beta}$, and the electrostatic energy of point-like charges is finite \cite{Kruglov1}.

Let us consider principles of causality and unitarity \cite{Shabad2}. They require the inequalities
\[
 {\cal L}_{\cal F}\leq 0,~~~~{\cal L}_{{\cal F}{\cal F}}\geq 0,
\]
\begin{equation}
{\cal L}_{\cal F}+2{\cal F} {\cal L}_{{\cal F}{\cal F}}\leq 0,
\label{2}
\end{equation}
where ${\cal L}_{\cal F}\equiv\partial{\cal L}/\partial{\cal F}$.
With the help of Eq. (1) one finds
\[
{\cal L}_{\cal F}= -\frac{1}{\left(1+\frac{4\beta{\cal F}}{3}\right)^{1/4}},~~~~ {\cal L}_{{\cal F}{\cal F}}=\frac{\beta}{3\left(1+\frac{4\beta{\cal F}}{3}\right)^{5/4}},
\]
\begin{equation}
{\cal L}_{\cal F}+2{\cal F} {\cal L}_{{\cal F}{\cal F}}=-\frac{3+2\beta{\cal F}}{3\left(1+\frac{4\beta{\cal F}}{3}\right)^{5/4}}.
\label{3}
\end{equation}
Thus, at $\beta{\cal F}\geq -3/2$ ($\beta{\cal F}\neq -3/4$) the principles of causality and unitarity hold.

The symmetric energy-momentum tensor corresponding to our model is
\begin{equation}
T_{\mu}^{~\nu}=-\frac{F_\mu^\alpha F_\alpha^\nu}{\left(1+\frac{4\beta{\cal F}}{3}\right)^{1/4}} -\delta^\nu_\mu{\cal L}.
\label{4}
\end{equation}
Making use of Eq. (4) we obtain the energy density
\begin{equation}
\rho=T_{t}^{~t}=\frac{E^2}{\left(1+\frac{4\beta{\cal F}}{3}\right)^{1/4}}- \frac{1}{\beta}\left[1-\left(1+\frac{4\beta{\cal F}}{3}\right)^{3/4}\right].
\label{5}
\end{equation}

\section{Dyonic solution}

BI-type-electrodynamics coupled to GR is described by the action
\begin{equation}
I=\int d^4x\sqrt{-g}\left(\frac{1}{16\pi G}R+ {\cal L}\right),
\label{6}
\end{equation}
with $G$ being the Newton constant. When $T_t^{~t}=T_r^{~r}$ the metric can be considered as static, spherically symmetric,
\begin{equation}
ds^2=-A(r)dt^2+\frac{1}{A(r)}dr^2+r^2(d\vartheta^2+\sin^2\vartheta d\phi^2).
\label{7}
\end{equation}
For the metric (7) nonzero components of $F^{\mu\nu}$ are $F^{tr}=-F^{rt}=E(r)$, $F^{\vartheta\phi}=-F^{\phi\vartheta}$, where $E(r)$ is the radial electric field.
Then field equations are given by
\begin{equation}
G_\mu^{~\nu}\equiv R_\mu^{~\nu}-\frac{1}{2}\delta_\mu^{~\nu}R=-8\pi GT_\mu^{~\nu},
\label{8}
\end{equation}
\begin{equation}
\partial_\mu\left(r^2{\cal L}_{{\cal F}}F^{\mu\nu}\right)=0.
\label{9}
\end{equation}
From Eq. (9)  we obtain \cite{Bronnikov3}, \cite{Bronnikov4}
\begin{equation}
E^2=\frac{q_e^2}{{\cal L}^2_{\cal F}r^4},
\label{10}
\end{equation}
where $q_e$ is the electric charge (the constant of integration).
From Bianchi identities $\nabla_\mu\tilde{F}^{\mu\nu}=0$ ($\tilde{F}^{\mu\nu}$ is the dual tensor) one finds
\begin{equation}
B^2=\frac{q^2_m}{r^4},
\label{11}
\end{equation}
with $q_m$ being the magnetic charge ($F^{\vartheta\phi}=q_m\sin\vartheta$).
With the help of Eqs. (10) and (11) we obtain
\begin{equation}
E^2=\frac{q_e^2}{r^4}\sqrt{1+\frac{4\beta{\cal F}}{3}},~~~~
{\cal F}=\frac{q_m^2}{2r^4}-\frac{q_e^2}{2r^4}\sqrt{1+\frac{4\beta{\cal F}}{3}}.
\label{12}
\end{equation}
 The solution to quadratic equation (12) is given by
\begin{equation}
\beta{\cal F}=a+\frac{2}{3}b^2-\frac{b}{3}\sqrt{12a+4b^2+9},~~~a=\frac{q_m^2\beta}{2r^4},
~~~b=\frac{q_e^2\beta}{2r^4},
\label{13}
\end{equation}
where the parameters $a$ and $b$ are unitless. Making use of Eq. (13) we find the dyonic solution for the electric field
\begin{equation}
E^2=\frac{2b}{3\beta}\left(\sqrt{12a+4b^2+9}-2b\right)=
\frac{q_e^2}{3r^4}\left(\sqrt{\frac{6q_m^2\beta}{r^4}+\frac{q_e^4\beta^2}{r^8}+9}-
\frac{q_e^2\beta}{r^4}\right).
\label{14}
\end{equation}
If $q_e=q_m$, we obtain from Eq. (14) Coulomb's law $E=q_e/r^2$. At the limit $r\rightarrow 0$ and when $q_m=0$ one finds from Eq. (14) $E_{max}= \sqrt{3}/\sqrt{2\beta}$.
At $r\rightarrow\infty$ we obtain from Eq. (14)
\begin{equation}
E=\frac{q_e}{r^2}+\frac{\beta q_e(q_m^2-q_e^2)}{6r^{6}}-\frac{\beta^2 q_e(q_m^2-q_e^2)(3q_m^2+q_e^2)}{72r^{10}}+{\cal O}(r^{-14}).
\label{15}
\end{equation}
Equation (15) shows the corrections to Coulomb's law. If $q_m=q_e$ corrections to Coulomb's law are absent.

The metric function $A(r)$ in Eq. (7) can be written as \cite{Bronnikov}
\begin{equation}
A(r) = 1-\frac{2M(r)G}{r}.
\label{16}
\end{equation}
Replacing Eq. (16) into Einstein equation (8), we obtain
\begin{equation}
G_t^{~t}=G_r^{~r}=-\frac{2M'(r)G}{r^2}=-8\pi GT_t^{~t}=-8\pi GT_r^{~r}=-8\pi G\rho,
\label{17}
\end{equation}
where the energy density $\rho$ is given by Eq. (5).
After the integration of Eq. (17), we find
\begin{equation}
M(r)=4\pi\int\rho r^2dr=m-4\pi\int_r^\infty\rho(r)r^2dr.
\label{18}
\end{equation}
The $m$ (the constant of integration) is the total mass of BH which can be considered as a free parameter.
Thus, Eqs. (16) and (18) are solutions of the Einstein equation. For the dyonic configuration, making use of Eqs. (5), (10), (11) and (14), we obtain the energy density
\begin{equation}
\beta\rho(r)=\frac{2a-2\beta{\cal F}}{\left(1+\frac{4\beta{\cal F}}{3}\right)^{1/4}}+ \left(1+\frac{4\beta{\cal F}}{3}\right)^{3/4}-1.
\label{19}
\end{equation}
Introducing the dimensionless variables $z=r^4/(\beta q_e^2)$, $n=q_m^2/q_e^2$ Eq. (19) is converted into the equation
\[
\beta\rho(z)=\frac{3z+2n-3zf(z)}{2zf^{1/4}(z)}+f^{3/4}(z)-1,
\]
\begin{equation}
f(z)=1+\frac{2}{3z}\left(n+\frac{1-\sqrt{6nz+9z^2+1}}{3z}\right).
\label{20}
\end{equation}
Making use of Eqs. (18) and (20) we find
\begin{equation}
M(z)=m-\frac{\pi q_e^{3/2}}{\beta^{1/4}}\int_z^\infty\biggl(\frac{3z+2n-3zf(z)}{2zf^{1/4}(z)}+f^{3/4}(z)-1
\biggr)\frac{dz}{z^{1/4}}.
\label{21}
\end{equation}
One may estimate the mass function at $z>1$ ($\beta<r^4/q_e^2$). At $z>1$ the energy density (20) becomes
\begin{equation}
\beta\rho(z)=\frac{1+n}{2z}-\frac{(n-1)^2}{24z^2}+\frac{(n-1)^2(5n+1)}{432z^3}+{\cal O}(z^{-4}).
\label{22}
\end{equation}
From Eq. (21) we obtain the mass function at $z>1$ ($\beta<r^4/q_e^2$)
\begin{equation}
M(z)=m+\frac{4\pi q_e^{3/2}}{\beta^{1/4}}\left(-\frac{n+1}{2z^{1/4}}+\frac{(n-1)^2}{120z^{5/4}}
-\frac{(n-1)^2(5n+1)}{3888z^{9/4}}\right)+{\cal O}(z^{-13/4}).
\label{23}
\end{equation}
Making use of Eqs. (16) and (23) one finds the metric function
\begin{equation}
A(z)=1-\frac{B}{z^{1/4}}+C\left(\frac{n+1}{\sqrt{z}}-\frac{(n-1)^2}{60z^{3/2}}
+\frac{(n-1)^2(5n+1)}{1944z^{5/2}}\right)+{\cal O}(z^{-7/2}),
\label{24}
\end{equation}
\begin{equation}
B=\frac{2mG}{\beta^{1/4}\sqrt{q_e}},~~~~C=\frac{4\pi q_eG}{\sqrt{\beta}},
\label{25}
\end{equation}
with the unitless variables $B$ and $C$. The authors \cite{Hendi1} also have considered corrections of Einstein-Maxwell BH and thermodynamics.
One can see the plots of the metric function for different values of $B$, $C$ and $n=q_m^2/q_e^2$ in Figs. (1) and (2).
\begin{figure}[h]
\includegraphics[height=3.0in,width=3.0in]{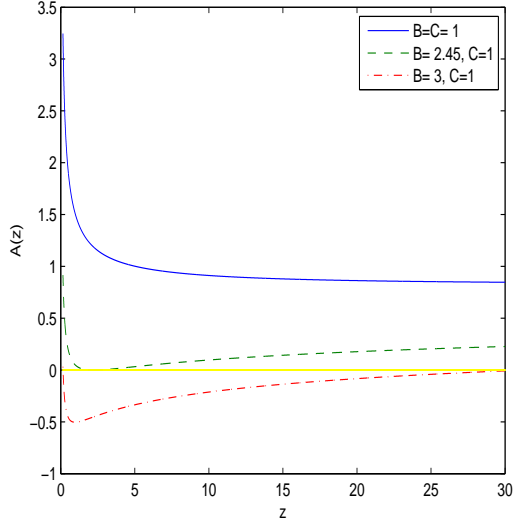}
\caption{\label{fig.1}The metric function $A(z)$ for $n=0.5$. Solid line corresponds to $B=C=1$, dashed line corresponds to $B=2.45$, $C=1$, and dashed-dotted line corresponds to $B=3$, $C=1$.}
\end{figure}
\begin{figure}[h]
\includegraphics[height=3.0in,width=3.0in]{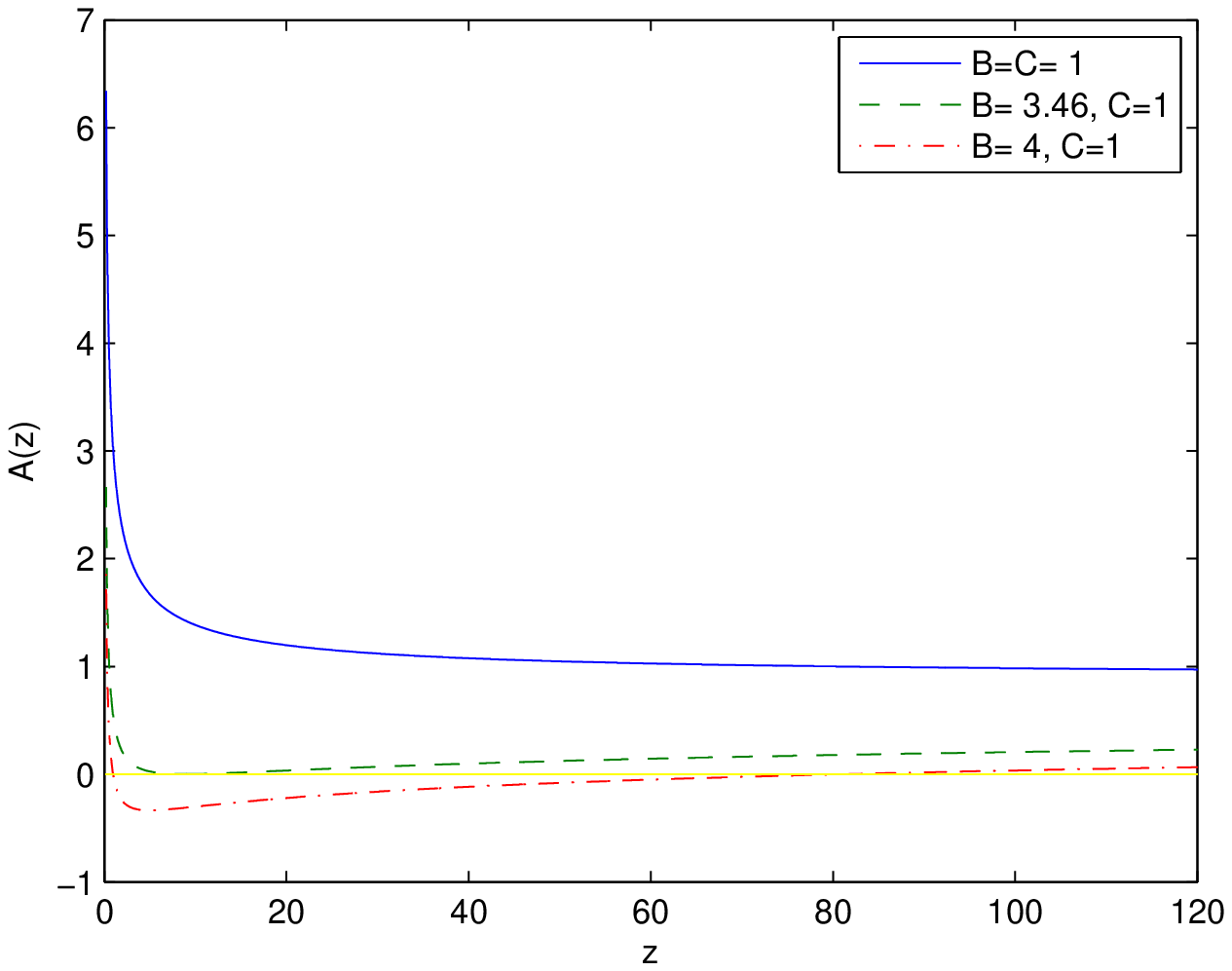}
\caption{\label{fig.2}The metric function $A(z)$ for $n=2$. Solid line corresponds to $B=C=1$, dashed line corresponds to $B=3.46$, $C=1$, and dashed-dotted line corresponds to $B=4$, $C=1$.}
\end{figure}
According to figures, at some parameters we have BH solutions with horizons, solutions with naked singularities, and extremal BH solutions. Fig. 1 shows that at $B=C=1$ ($q_e^2=2q^2_m$) the naked singularity occurs, at $B=2.45$, $C=1$ ($q_e^2=2q_m^2$) an extremal BH solution holds, and at $B=3$, $C=1$ ($q_e^2=2q^2_m$) there is the BH solution with two horizons. In accordance with Fig. 2 the similar situation for $q_e^2=0.5q^2_m$ occurs. For $q_m>q_e$ the event horizon of BH is bigger comparing to the case $q_e>q_m$.

Taking into account the notations $z=r^4/(\beta q_e^2)$, $n=q_m^2/q_e^2$, the metric function (24) as $\beta<r^4/q_e^2$ is given by
\[
A(r)=1-\frac{2mG}{r}+\frac{4\pi(q_e^2+q_m^2)G}{r^2}-\frac{\pi\beta(q_m^2-q_e^2)^2G}{15r^6}
\]
\begin{equation}
+\frac{\pi\beta^2(q_m^2-q_e^2)^2(5q_m^2+q_e^2)G}{486r^{10}}
+{\cal O}(\beta^3).
\label{26}
\end{equation}
According to Eq. (26) there are corrections to the Reissner$-$Nordstr\"{o}m solution. From Eq. (20) at $q_e=q_m$ ($n=1$) we have $\beta\rho(z)=1/z$ and corrections to the Reissner$-$Nordstr\"{o}m solution vanish.

\section{Thermodynamics}

The Hawking temperature reads
\begin{equation}
T_H=\frac{\kappa}{2\pi}=\frac{A'(r_+)}{4\pi},
\label{27}
\end{equation}
were $\kappa$ is the surface gravity and $r_+$ is the event horizon. With the help of Eqs. (16) and (18) one obtains
\begin{equation}
A'(r)=\frac{2 GM(r)}{r^2}-\frac{2GM'(r)}{r},~~~M'(r)=4\pi r^2\rho(r),~~~M(r_+)=\frac{r_+}{2G}.
\label{28}
\end{equation}
Making use of Eqs. (27) and (28) we find
\begin{equation}
T_H(r_+)=\frac{1}{4\pi}\left(\frac{1}{r_+}-8\pi Gr_+\rho(r_+)\right).
\label{29}
\end{equation}
From Eqs. (20), (28)  and (29) one obtains the Hawking temperature
\begin{equation}
T_H(z_+)=\frac{1}{4\pi\beta^{1/4}\sqrt{q_e}}\biggl[\frac{1}{z^{1/4}_+}-\frac{8\pi Gq_ez_+^{1/4}}{\sqrt{\beta}}
\left(\frac{2n+3z_+(1-f(z_+))}{2z_+f^{1/4}(z_+)}+f^{3/4}(z_+)-1\right)\biggr],
\label{30}
\end{equation}
with $z_+=r_+^4/(\beta q_e^2)$.
From the relation $2GM(r_+)=r_+$ we see that the event horizon $r_+$ (and $z_+$) and the parameter $\beta$ obey
\begin{equation}
\frac{2Gq_e}{\sqrt{\beta}}=\frac{z_+^{1/4}}{\frac{m\beta^{1/4}}{q_e^{3/2}}-\pi\int_{z_+}^\infty
\left(\frac{2n+3z_+(1-f(z_+))}{2z_+f^{1/4}(z_+)}+f^{3/4}(z_+)-1\right)\frac{dz}{z^{1/4}}}.
\label{31}
\end{equation}
Placing unitless variable $2Gq_e/\sqrt{\beta}$ from Eq. (31) into Eq. (30) we obtain the Hawking temperature of BH
\[
T_H(z_+)=\frac{1}{4\pi\beta^{1/4}\sqrt{q_e}}\biggl(\frac{1}{z^{1/4}_+}
\]
\begin{equation}
-\frac{4\pi\sqrt{z_+}\left(\frac{2n+3z_+(1-f(z_+))}{2z_+f^{1/4}(z_+)}+f^{3/4}(z_+)-1\right)}
{\frac{m\beta^{1/4}}{q_e^{3/2}}-
\pi\int_{z_+}^\infty\left(\frac{2n+3z(1-f(z))}{2zf^{1/4}(z)}+f^{3/4}(z)-1\right)
\frac{dz}{z^{1/4}}}\biggr).
\label{32}
\end{equation}
If $z_+>1$ we can use the approximate value for the integral in Eq. (32) and obtain
\[
T_H(z_+)=\frac{1}{4\pi\beta^{1/4}\sqrt{q_e}}\biggl(\frac{1}{z^{1/4}_+}
\]
\begin{equation}
-\frac{\sqrt{z_+}\left(\frac{2n+3z_+(1-f(z_+))}{2z_+f^{1/4}(z_+)}+f^{3/4}(z_+)-1\right)}
{P-\frac{n+1}{2z_+^{1/4}}+\frac{(n-1)^2}{120z_+^{5/4}}-\frac{(n-1)^2(5n+1)}{3888z_+^{9/4}}+{\cal O}(z_+^{13/4})}
\biggr),~~~P=\frac{m\beta^{1/4}}{4\pi q_e^{3/2}}.
\label{33}
\end{equation}
We use series for the energy density $\rho$ in the parameter $1/z_+=\beta q_e^2/r_+^4 <1$ which can be considered as series in the small parameter $\beta<r_+^4/q_e^2$ and, therefore, the Hawking temperature is valid for the horizon of the black holes. For positive heat capacity BHs are stable and otherwise they are unstable. We will investigate phase transitions by considering the heat capacity. When the heat capacity is singular the second-order phase transition occurs. From the  Hawking entropy of BH $S=\mbox{Area}/(4G)=\pi r_+^2/G$ we obtain the heat capacity
\begin{equation}
C_q=T_H\left(\frac{\partial S}{\partial T_H}\right)_q=\frac{T_H\partial S/\partial r_+}{\partial T_H/\partial r_+}=\frac{2\pi r_+T_H}{G\partial T_H/\partial r_+}.
\label{34}
\end{equation}
In accordance with Eq. (34) the heat capacity diverges if the Hawking temperature has the extremum, $\partial T_H/\partial r_+=0$.
In Figs. 3 and 4 the plots of Hawking temperature versus the variable $z_+^{1/4}=r_+/(\beta^{1/4}\sqrt{q_e})$ for different parameters $P$ and $n$ are depicted.
\begin{figure}[h]
\includegraphics[height=3.0in,width=3.0in]{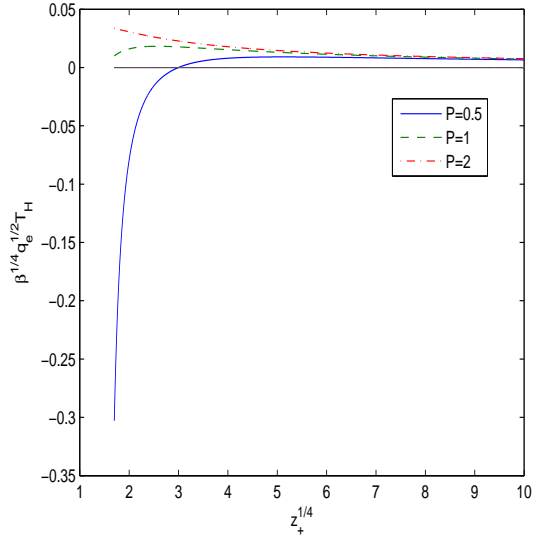}
\caption{\label{fig.3}The Hawking temperature vs. $z_+^{1/4}=r_+/(\beta^{1/4}\sqrt{q_e}) $ for $n=0.5$. Solid line corresponds to $P=0.5$, dashed line corresponds to $P=1$, and dashed-dotted line corresponds to $P=2$.}
\end{figure}
\begin{figure}[h]
\includegraphics[height=3.0in,width=3.0in]{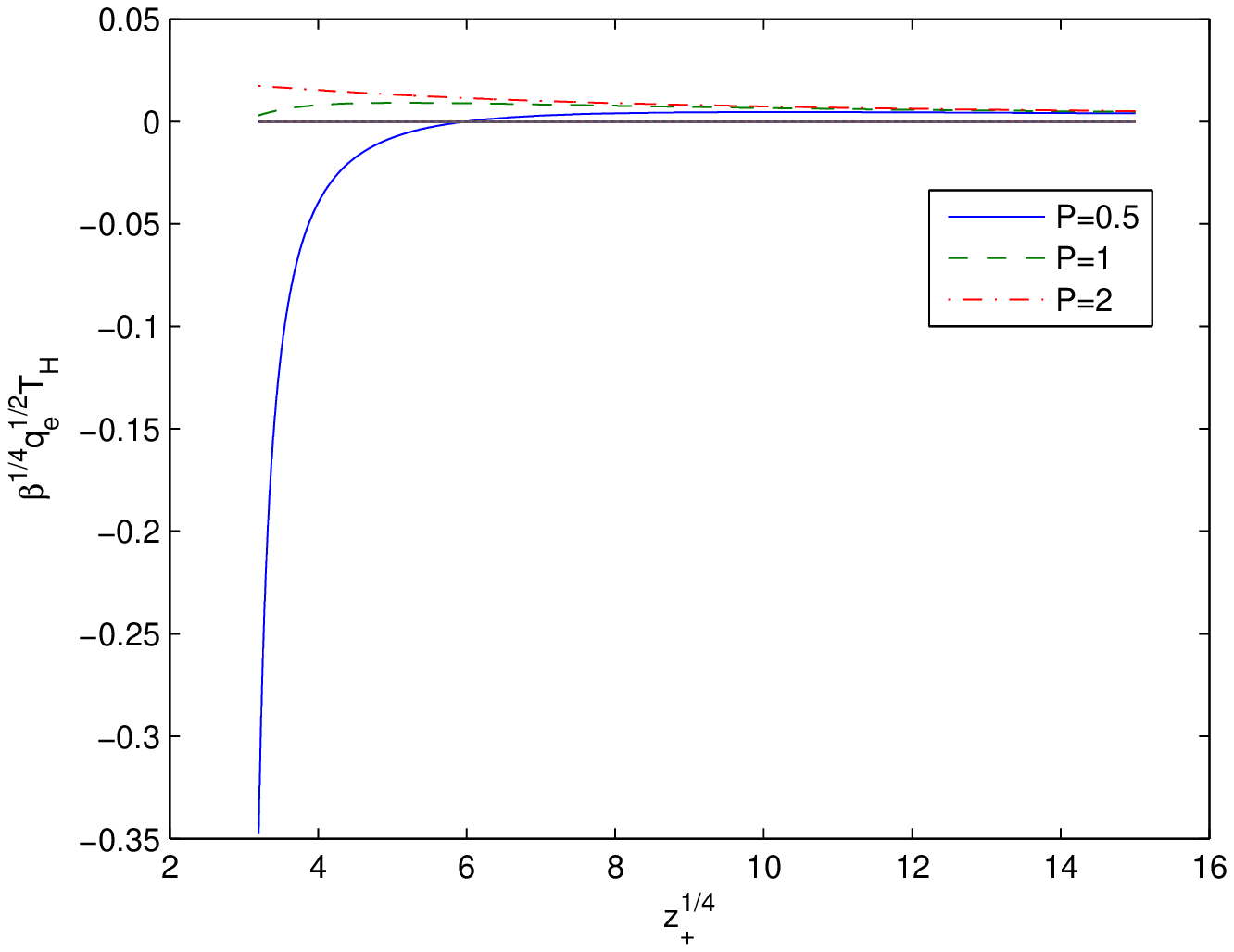}
\caption{\label{fig.4}The Hawking temperature vs. $z_+^{1/4}=r_+/(\beta^{1/4}\sqrt{q_e})$ for $n=2$. Solid line corresponds to $P=0.5$, dashed line corresponds to $P=1$, and dashed-dotted line corresponds to $P=2$.}
\end{figure}
 Fig. 3 shows that at $n=0.5$ more massive BHs exist (with the positive Hawking temperature) for smaller values of $r_+$ ($z_+$) and the Hawking temperature has the extremum.
 According to Fig. 4 similar behavior takes place when $n=2$ for large $r_+$.
 If Hawking temperatures are negative, one can say that BHs are absent. Figs. 5 and 6 show the plots of the heat capacity versus the variable $z_+^{1/4}=r_+/(\beta^{1/4}\sqrt{q_e)}$ for different parameters $P$ and $n$.
  \begin{figure}[h]
\includegraphics[height=3.0in,width=3.0in]{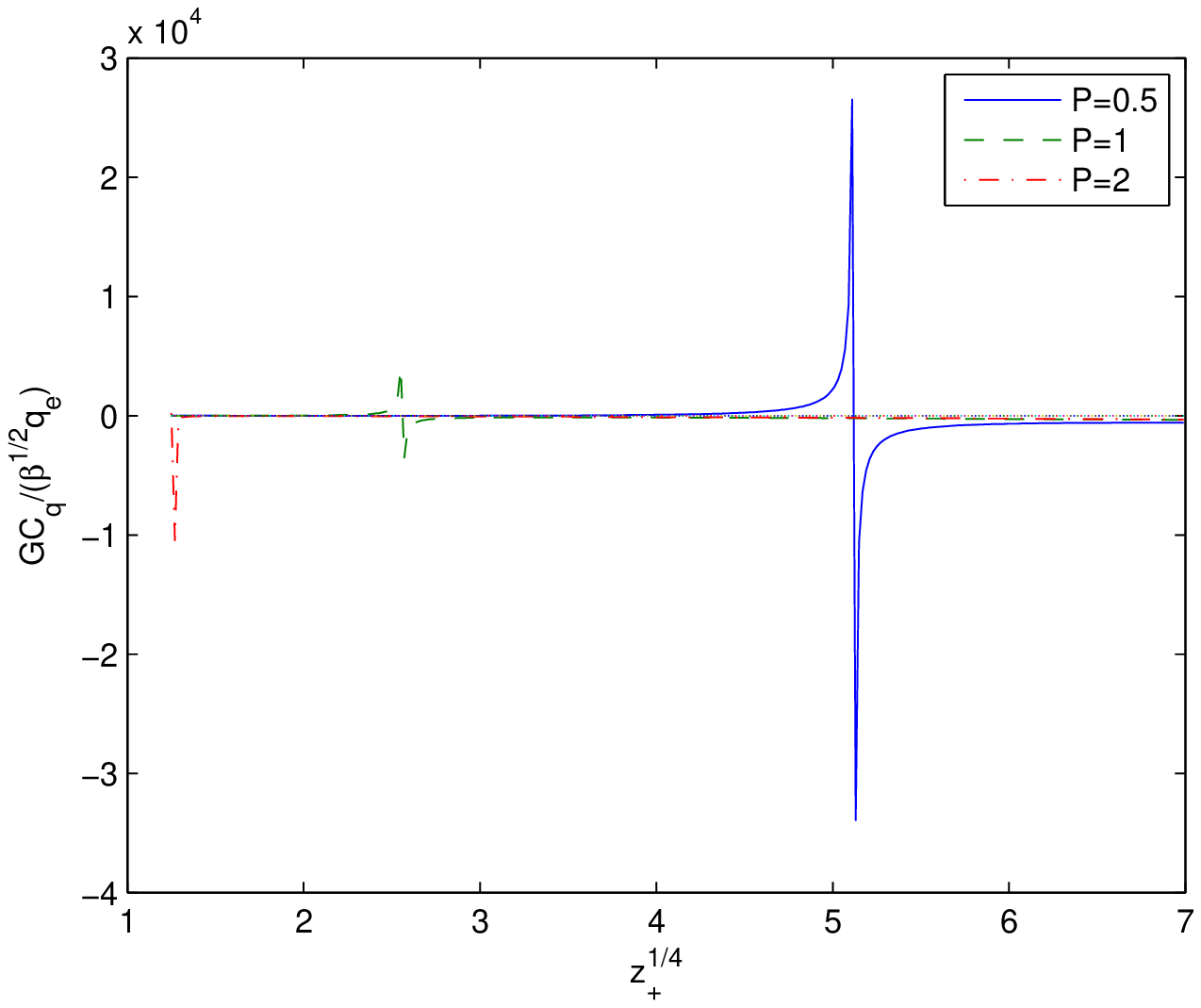}
\caption{\label{fig.5}The heat capacity vs. $z_+^{1/4}=r_+/(\beta^{1/4}\sqrt{q_e}) $ for $n=0.5$. Solid line corresponds to $P=0.5$, dashed line corresponds to $P=1$, and dashed-dotted line corresponds to $P=2$.}
\end{figure}
\begin{figure}[h]
\includegraphics[height=3.0in,width=3.0in]{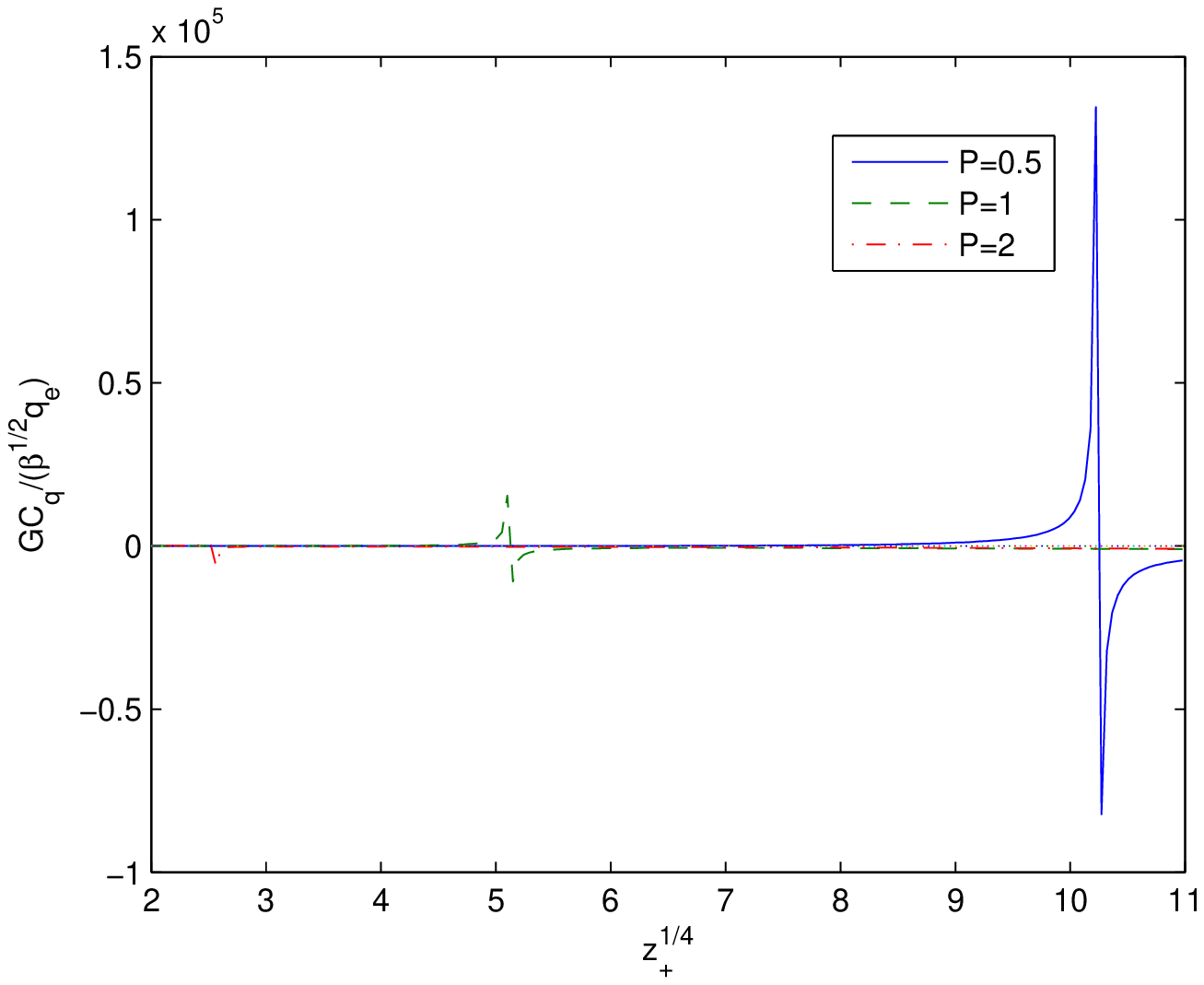}
\caption{\label{fig.6}The heat capacity vs. $z_+^{1/4}=r_+/(\beta^{1/4}\sqrt{q_e})$ for $n=2$. Solid line corresponds to $P=0.5$, dashed line corresponds to $P=1$, and dashed-dotted line corresponds to $P=2$.}
\end{figure}
Figures 5 and 6 show the phase transitions in the points (the event horizons) where the heat capacities possess singularities. These points correspond to the extremum of the Hawking temperatures depicted in figures 3 and 4.
In accordance with Fig. 5 in the case $n=0.5$ ($q_e^2=2q_m^2$), for massive BH's the phase transitions take place for smaller values of event horizons ($r_+$). Fig. 6 shows similar behavior of the heat capacity for the case $n=2$ ($q_m^2=2q_e^2$) but event horizons ($r_+$), for the phase transitions, are greater. One can see from Figs. 5 and 6 that there is a range for $r_+$ where BHs are stable. This is in agreement with the work \cite{Hendi1}.

\section{Conclusion}

BI-type electrodynamics with the parameter $\beta$ coupled with GR have been studied. We have obtained the dyonic BH solution. The correspondence principle in this model holds so that at the weak field limit we have the Maxwell's Lagrangian. The corrections to Coulomb's law and Reissner$-$Nordstr\"{o}m solution at $\beta<r^4/q_e^2$ were  found. We demonstrated that for self-duality case at $q_e=q_m$ corrections are absent. The Hawking temperature and heat capacity of BH were calculated and we shown that phase transitions, within our model, occur for some model parameters. We shown that there is a range for $r_+$, depending on the BH mass and $n$, where BHs are stable (the heat capacity is positive).

\section{Appendix: The Kretschmann scalar}

 BH singularities can be investigated by considering the Kretschmann scalar $K(r)$  which is given by \cite{Hendi1}
\begin{equation}\label{35}
 K(r)\equiv R_{\mu\nu\alpha\beta}R^{\mu\nu\alpha\beta}=A''^2(r)+\left(\frac{2A'(r)}{r}\right)^2
+\left(\frac{2A(r)}{r^2}\right)^2,
\end{equation}
and $A'(r)=\partial A(r)/\partial r$. For dyonic configuration the metric function $A(r)$ at $r^4>\beta q_e^2$ is given by Eq. (24). Making use of Eqs. (26) and (35) we obtain
\begin{equation}\label{36}
  \lim_{r\rightarrow\infty} K(r)=0.
\end{equation}
Thus, in accordance with Eq. (36) spacetime is flat at $r\rightarrow\infty$ and there no singularity.
One can see from Eqs. (16) and(21) that the metric function at $r\rightarrow 0$ becomes singular.
As a result, according to Eq. (35) we have
\begin{equation}\label{37}
 \lim_{r\rightarrow\ 0} K(r)=\infty,
\end{equation}
and the Kretschmann scalar possesses the singularity at $r=0$ as well as in other NED models \cite{Hendi1}.

\end{document}